\documentclass[aps,pra,epsfig,twocolumn,showpacs,superscriptaddress]{revtex4}

\def\be{\begin{equation}}
\def\ee{\end{equation}}
\def\bea{\begin{eqnarray}}
\def\eea{\end{eqnarray}}
\def\bma{\begin{mathletters}}
\def\ema{\end{mathletters}}
\def\bi{\begin{itemize}}
\def\ei{\end{itemize}}

\def\C{\hbox{$\mit I$\kern-.7em$\mit C$}}

\begin{document}

\title{Probabilistic cloning and signalling}

\author{Sibasish Ghosh}
\email{sibasish@imsc.res.in}
\affiliation{Institute of Mathematical Sciences,
C. I. T. Campus, Taramani, Chennai 600113, India}

\author{Guruprasad Kar}
\email{gkar@imsc.res.in}
\affiliation{Institute of Mathematical Sciences,
C. I. T. Campus, Taramani, Chennai 600113, India}

\author{Samir Kunkri}
\email{skunkri_r@isical.ac.in}
\affiliation{Physics and Applied Mathematics Unit,
Indian Statistical Institute, 203 B. T. Road, Kolkata 700108, India}

\author{Anirban Roy}
\email{anirb@imsc.res.in}
\affiliation{Institute of Mathematical Sciences,
C. I. T. Campus, Taramani, Chennai 600113, India}

\begin{abstract}
We give a proof of impossibility of probabilistic exact
$1\rightarrow 2$ cloning of any three different states of a qubit.
The simplicity of the proof is due to the use of a surprising result
of remote state preparation [M. -Yong Ye, Y. -Sheng Zhang and G. -Can Guo,
quant-ph/0307027 (2003)]. The result is extented to higher dimentional
cases for special ensemble of states.
\end{abstract}


\maketitle

%\vspace{0.4cm}

%\doublespacing

%\vspace{0.4cm}
An arbitrary quantum state can not be cloned exactly because of
the no-cloning theorem \cite{wootters82}. However Duan and Guo
showed that probabilistic exact cloning for the set of linearly
independent  state is possible \cite{duan98}. It is known that if
the quantum states can be cloned exactly then physical massage can
be sent superluminaly. Hardy and Song showed that if probabilistic
exact cloning of $(d+1)$ number of quantum states, in which any
$d$ number of states are linearly independent, is possible, then
there will be signalling \cite{hardy99}. Pati showed that
probabilistic exact cloning of four states $|\psi\rangle,
|{\psi}^{\bot}\rangle, |\phi\rangle, |{\phi}^{\bot}\rangle$ of a
two dimensional Hilbert space implies (probabilistic) signalling,
as distinguishability of the two mixtures $\frac{1}{2}
(P[|\psi\rangle \otimes |\psi\rangle] + P[|{\psi}^{\bot}\rangle
\otimes |{\psi}^{\bot}\rangle])$ and  $\frac{1}{2} (P[|\phi\rangle
\otimes |\phi\rangle] + P[|{\phi}^{\bot}\rangle \otimes
|{\phi}^{\bot}\rangle])$ is probabilistically possible
\cite{pati00}.

In this paper we show that probabilistic exact
cloning of any three different states of a qubit
implies (probabilistic) {\it signalling } in the sense,
that one can extract more than 1 cbit message
probabilistically by communicating 1 cbit only \cite{ft}.
Here we use the technique of remote state preparation to
provide an alternative as well as simpler proof, in the
qubit case, given by Hardy and Song \cite{hardy99}.
We generalize this result in $d$ dimentional Hilbert
space, where we show that the probabilistic exact
cloning of $(d+1)$ number of states, in which $d$
number of states are linearly independent, taken
from a special ensemble of states, implies signalling.

It is an interesting property of ${C\!\!\!\!I}^2$ that
the Bloch vectors corresponding to any three different
states of a qubit say,
$|{\psi}_{1}\rangle,|{\psi}_{2}\rangle,|{\psi}_{3}\rangle $,
lie either on a great circle or a  small circle of the Bloch
sphere. Here it is to be mentioned that a small circle is
defined as circle formed by intersection of any non-diametral
plane and the Bloch sphere. Hence, given any three different
states
$|{\psi}_{1}\rangle,|{\psi}_{2}\rangle,|{\psi}_{3}\rangle $ of ${C\!\!\!\!I}^2$,
one can find an orthogonal basis $\{|0\rangle, |1\rangle\}$
of ${C\!\!\!\!I}^2$ (say), two positive numbers $\alpha_0$,
$\alpha_1$ (say) with ${\alpha}^2_0 +{\alpha}^2_1 = 1$, and
real numbers $\phi_{01}$, $\phi_{11}$, $\phi_{02}$, $\phi_{12}$,
$\phi_{03}$, $\phi_{13}$, such that the above-mentioned three
states can be expressed as
$|{\psi}_{k}\rangle = \sum_{j = 0}^{1} {\alpha}_{j} {e}^{i{\phi}_{jk}} |j\rangle$,
for $k = 1, 2, 3$.
%  As the states $|{\psi}_{1}\rangle,|{\psi}_{2}\rangle,|{\psi}_{3}\rangle $  are linearly dependent, therefore, one can not distinguish the states $|{\psi}_{1}\rangle,|{\psi}_{2}\rangle,|{\psi}_{3}\rangle $ unambiguously,  when only single copy of each of the states is provided. We know that collection of two copies of all state of ${C\!\!\!\!I}^2$  spans a three dimentional subspace of total four dimentional Hilbert space  ${C\!\!\!\!I}^2\otimes {C\!\!\!\!I}^2$.
As  $|{\psi}_{1}\rangle,|{\psi}_{2}\rangle,|{\psi}_{3}\rangle $
are different, therefore two copies of these three states, i.e.,
${|{\psi}_{1}\rangle}^{\otimes 2},{|{\psi}_{2}\rangle}^{\otimes 2},
{|{\psi}_{3}\rangle}^{\otimes 2}$ of ${C\!\!\!\!I}^2\otimes {C\!\!\!\!I}^2$,
are linearly independent. Then one can construct a POVM by which one can
distinguish any state unambiguously from these set of three states with
non-zero probability( less than one) \cite{peres98}.

Now Alice wants to prepare one of the three different states
$|{\psi}_{1}\rangle,|{\psi}_{2}\rangle,|{\psi}_{3}\rangle $,
which is the element of the ensemble, remotely at Bob's place,
in which she encoded three different messages. So these three
states can be expressed as
$|{\psi}_{k}\rangle = \sum_{j = 0}^{1} {\alpha}_{j}
{e}^{i{\phi}_{jk}} |j\rangle$, for $k = 1, 2, 3$. Alice can
perform the remote state preparation by using an entangled state,
$|\psi\rangle_{AB} = \sum_{i = 0}^{1} {\alpha}_{i}
|i\rangle_A \otimes |i\rangle_B$, shared between Alice and Bob,
and communicating 1 cbit only \cite{{patire},{yong03}}. Here
the orthogonal basis $\{|0\rangle, |1\rangle\}$ is same as that
used in the above-mentioned expressions for $|{\psi}_{k}\rangle$'s. We now assume that Bob has a $1\rightarrow 2$ probabilistic
quantum cloning machine (PQCM) by which he can exactly clone these
three linearly dependent states probabilistically. Bob will apply
his $1\rightarrow 2$ PQCM on his particle, after Alice prepares
the state at his place. Now we consider the case when Bob wll be
successful to make the exact clone of the state of his particle.
In this case, the state of Bob will be one of the three states of
${|{\psi}_{1}\rangle}^{\otimes 2},{|{\psi}_{2}\rangle}^{\otimes
2}, {|{\psi}_{3}\rangle}^{\otimes 2} $. Since these states are
linearly independent, Bob can distinguish these states
${|{\psi}_{1}\rangle}^{\otimes 2},{|{\psi}_{2}\rangle}^{\otimes
2}, {|{\psi}_{3}\rangle}^{\otimes 2} $ probabilistically which, in
turn, implies that Bob can probabilistically extract more than 1
cbit of classical information ({\it i.e}, decoding the above
mentioned three messages) probabilistically, although Alice has
spent 1 cbit during the remote state preparation. This implies
(probabilistic) signalling. Thus we conclude that probabilistic
exact cloning of
linearly dependent states from ${C\!\!\!\!I}^2$ is not possible.

Now we extend our argument in general $d$ dimensional Hilbert space,
for the special kinds of ensemble. For a given vector
${\vec{\alpha}} = ( {\alpha}_{0},{\alpha}_{1}, \dots,{\alpha}_{d-1})$
where $ {\alpha}_{i}> 0 $, $ \sum_{i = 0}^{d - 1} {{\alpha}_{i}}^{2} = 1$,
we choose the ensemble as 
\begin{equation}
\label{ensemble}
S_{\vec{\alpha}} = \left\{\sum_{j = 0}^{d - 1}
{\alpha}_{j} {e}^{i{\phi}_{j}} |j\rangle  :
( {\phi}_{0},  {\phi}_{1}, \ldots, {\phi}_{d-1}) \in {T}^{d}\right\},
\end{equation}
where ${T}^{d} \equiv T\times T\times \ldots d$ times, and
$T = \{ x \in {I\!\!R} : 0 \le x \le 2\pi\}$.\\
First of all we check whether we can get $d$ number of linearly
independent states from this ensemble. To do this, let us first
consider the case for $d = 3$. We choose three states from the
ensemble of equation (\ref {ensemble}) for $d = 3$. They are
\begin{equation}
\label{phases}
\begin{array}{lcl}
|{\psi}_{0}\rangle &=& {\alpha}_{0} |0\rangle + {\alpha}_{1} |1\rangle + {\alpha}_{2} |2\rangle, \\
|{\psi}_{1}\rangle &=& {\alpha}_{0} {e}^{i{\phi}_{0}}|0\rangle + {\alpha}_{1} {e}^{i{\phi}_{1}} |1\rangle + {\alpha}_{2} {e}^{i{\phi}_{2}} |2\rangle, \\
|{\psi}_{2}\rangle &=& {\alpha}_{0} {e}^{i{\phi}_{0}^{\prime}}|0\rangle + {\alpha}_{1} {e}^{i{\phi}_{1}^{\prime}}|1\rangle + {\alpha}_{2} {e}^{i{\phi}_{2}^{\prime}} |2\rangle.
\end{array}
\end{equation}
 The states $ |{\psi}_{0}\rangle, |{\psi}_{1}\rangle,
 |{\psi}_{2}\rangle $ are linearly independent iff
\begin{equation}
\label{independent}
{\lambda}_{0}|{\psi}_{0}\rangle + {\lambda}_{1}|{\psi}_{1}\rangle
+ {\lambda}_{2}|{\psi}_{2}\rangle = 0
\end{equation}
when and only when
$ ({\lambda}_{0}, {\lambda}_{1}, {\lambda}_{2}) = (0, 0, 0) $.
From equation  (\ref {phases}) and  (\ref {independent}) we get
\begin{equation}
\label{coefficient}
\begin{array}{lcl}
{\lambda}_{0} + {\lambda}_{1} {e}^{i{\phi}_{0}} + {\lambda}_{2} {e}^{i{\phi}_{0}^{\prime}} &=& 0, \\
{\lambda}_{0} + {\lambda}_{1} {e}^{i{\phi}_{1}} + {\lambda}_{2} {e}^{i{\phi}_{1}^{\prime}} &=& 0, \\
{\lambda}_{0} + {\lambda}_{1} {e}^{i{\phi}_{2}} + {\lambda}_{2} {e}^{i{\phi}_{2}^{\prime}} &=& 0 .
\end{array}
\end{equation}
Thus the three states
$ |{\psi}_{0}\rangle, |{\psi}_{1}\rangle, |{\psi}_{2}\rangle $
will be linearly independent iff the determinant of the
coefficient matrix of equation  (\ref {coefficient})  must
be non-zero, {\it i.e.},
\begin{equation}
\label{determinant}
\left|
\begin{array}{ccc}
1 &{e}^{i{\phi}_{0}}& {e}^{i{\phi}_{0}^{\prime}} \\
1 &{e}^{i{\phi}_{1}}& {e}^{i{\phi}_{1}^{\prime}} \\
1 &{e}^{i{\phi}_{2}}& {e}^{i{\phi}_{2}^{\prime}}.
\end{array}
\right| \ne 0.
\end{equation}
{\it i.e.},
$$ {e}^{i({\phi}_{1} + {\phi}_{2}^{\prime})} -
{e}^{i({\phi}_{1} + {\phi}_{0}^{\prime})} -
{e}^{i({\phi}_{0} + {\phi}_{2}^{\prime})}$$
%\begin{widetext}
\begin{equation}
\label{determinant1}
- {e}^{i({\phi}_{2} + {\phi}_{1}^{\prime})}
+ {e}^{i({\phi}_{0} + {\phi}_{1}^{\prime})} +
{e}^{i({\phi}_{2} + {\phi}_{0}^{\prime})} \ne 0.
%\end{array}
\end{equation}
%\end{widetext}

So one can always choose the phases ${\phi}_{0},  {\phi}_{1},
{\phi}_{2}, {\phi}_{0}^{\prime}, {\phi}_{1}^{\prime},
{\phi}_{2}^{\prime}$, such that it will satisfy the equation (\ref
{determinant1}). Thus the states $|{\psi}_{0}\rangle$,
$|{\psi}_{1}\rangle$, $|{\psi}_{2}\rangle$ are linearly
independent for the correct choice of the phases satisfying
equation (\ref {determinant1}). We can extend our argument for
any arbitrary $d$ dimensional Hilbert space. So we can conclude
that one can always choose $d$ number of linearly independent
states from the ensemble of states given in equation (\ref{ensemble}).

Let us now choose $(d+1)$ number of states from the ensemble of
states given in equation (\ref{ensemble}), in which $d$ number
of states are linearly independent. We now show that the two
copies of each of these  $(d + 1)$ number of states are linearly
independent. To prove this again we consider the case for $d=3$.
We choose four different states  $|{\psi}_{0}\rangle$,
$|{\psi}_{1}\rangle$, $|{\psi}_{2}\rangle$ and  $|{\psi}_{3}\rangle$
from the ensemble of state given in equation (\ref{ensemble})
for $d=3$ in which $|{\psi}_{0}\rangle$, $|{\psi}_{1}\rangle$,
$|{\psi}_{2}\rangle$ are linearly independent. Let we assume
that two copy of each of these four state are not linearly
independent. Then the state ${|{\psi}_{3}\rangle}^{\otimes 2}$
can be written as the linear combination of the three states
${|{\psi}_{0}\rangle}^{\otimes 2},
{|{\psi}_{1}\rangle}^{\otimes 2},{|{\psi}_{2}\rangle}^{\otimes 2}$,
{\it i.e.},
\begin{equation}
\label{linear}
{|{\psi}_{3}\rangle}^{\otimes 2} =
{\lambda}_{0}{|{\psi}_{0}\rangle}^{\otimes 2}
+ {\lambda}_{1}{|{\psi}_{1}\rangle}^{\otimes 2}
+ {\lambda}_{2}{|{\psi}_{2}\rangle}^{\otimes 2}
\end{equation}
L.H.S of the equation (\ref{linear}) is a product
state but R.H.S is an entangle state, because
$ {|{\psi}_{0}\rangle}^{\otimes 2},
{|{\psi}_{1}\rangle}^{\otimes 2},
{|{\psi}_{2}\rangle}^{\otimes 2}$ are
linearly independent and at least two of
${\lambda}_{0}, {\lambda}_{1}, {\lambda}_{2}$ are
non-zero.  Thus the state $ {|{\psi}_{3}\rangle}^{\otimes 2}$
can not be written as the linear combination of of the states
$ {|{\psi}_{0}\rangle}^{\otimes 2},
{|{\psi}_{1}\rangle}^{\otimes 2},{|{\psi}_{2}\rangle}^{\otimes 2}$.
Then the two copies of each of the states  $|{\psi}_{0}\rangle$,
$|{\psi}_{1}\rangle$, $|{\psi}_{2}\rangle, |{\psi}_{3}\rangle$
are linearly independent. We can extend our argument for $d$
dimensional Hilbert space. Thus we can conclude that two copies
each of these $ (d + 1)$ number of states are linearly independent.

In order to extend our argument for general $d$ dimension,
let us assume that, Alice and Bob share an entangle state
in two $d$-dimensional systems
\begin{equation}
\label{entangle}
|\psi\rangle_{AB} = \sum_{i = 0}^{d - 1} {\alpha}_{i} |i\rangle_A \otimes |i\rangle_B,
\end{equation}
where ${\alpha}_{i} > 0$, and $\sum_{i = 0}^{d - 1}
{\alpha}_{i}^{2} = 1$. Alice can remotely prepare states from  any
given subset from the ensemble in equation (\ref{ensemble}) by
using the entangled state given in equation (\ref{entangle}) and
communicating ${\rm log}_2 d$ cbits only\cite{yong03}. Alice wants
to prepare remotely (at Bob's place) one of the $(d+1)$ number of
states in which $d$ number of states are linearly independent,
chosen from the ensemble given in equation (\ref{ensemble}). Now
we assume that Bob has a $1\rightarrow 2$  PQCM which can clone
these $(d + 1)$ number of states exactly.  Bob will apply his
$1\rightarrow 2$ PQCM on his state after Alice prepare the state
to him. Now we consider the case when Bob wll be successful to
make the exact clone of his state. Since two copies of each of
these $(d + 1)$ states are linearly independent ( which we prove
earlier), then Bob can distinguish his state probabilistically. So
probabilistically Bob can extract more than  ${\rm log}_2 d$ cbits
of information from Alice's sent message (${\rm log}_2 d$ cbits of
information), by using his PQCM, which implies probabilistic
signalling. Thus we can conclude that probabilistic exact cloning
of linearly dependent states from the ensemble given in equation
(\ref{ensemble}), implies (probabilistic) signalling.

Our argument would run for a most general set of $(d + 1)$ number
of linearly dependent states, among which any $d$ number of states
are linearly independent (which we denote by
$|{\chi}_{1}\rangle,|{\chi}_{2}\rangle,\dots|{\chi}_{d}\rangle $),
if we could have found an unitary operator $U$, which takes
$|{\chi}_{k}\rangle$ to an element $\sum_{j = 0}^{d - 1}
{\alpha}_{j} {e}^{i{\phi}_{jk}} |j\rangle $ of $S_{\vec{\alpha}}$
(for given ${\vec{\alpha}}$) for $k = 1, 2, \dots (d+1)$, when
$|{\chi}_{(d+1)}\rangle$ is the $(d + 1)$th element of the
above-mentioned linearly dependent set
$|{\chi}_{1}\rangle,|{\chi}_{2}\rangle,\dots|{\chi}_{d+1}\rangle $
\cite{ft1}. This means that if the $(d+1)$th state is given by
$|{\chi}_{d+1}\rangle = \sum_{k = 1}^{d} {\lambda}_{k}
|{\chi}_k\rangle $, ( $\lambda_{k} \in {C\!\!\!\!I}$ ), then
$\sum_{j = 0}^{d - 1} {\alpha}_{j}\{ \sum_{k = 1}^{d}
{\lambda}_{k}  {e}^{i{\phi}_{jk}}\} |j\rangle $ must be an element
of $S_{\vec{\alpha}}$, which, in turn implies that $ \sum_{k =
1}^{d} {\lambda}_{k}  {e}^{i{\phi}_{jk}} =  {e}^{i{\theta}_{j}}$
for $j = 1, 2, \dots d $ and $ {\theta}_{j}$'s are real numbers.
This, in general, does not hold good for arbitrary choice of
$|{\chi}_{1}\rangle,|{\chi}_{2}\rangle,\dots|{\chi}_{d+1}\rangle
$, even if such an $U$ would
 exit.

 Recently Zeng and Zhang \cite{zeng} have shown that Alice can
 remotely (exactly) prepare any state of the form $|{\chi}\rangle =
 a_0|0\rangle + a_1|1\rangle + \dots + a_{(d-1)}|d-1\rangle $ (
 where $a_i$'s are all real coefficients, $\sum_{i = 0}^{d-1} a_i^2 = 1$, and $\{|0\rangle, |1\rangle, \dots |d-1\rangle\}$
 is a fixed orthonormal basis of a $d$ dimensional Hilbert
 space) at Bob's place, using ${\rm log}_2 d$ amount of shared
 free ebit and ${\rm log}_2 d$ amount of classical communication
 from Alice to Bob, if and only if $ d = 2$, or $4$, or $8$. So let us
 take $d$ to be either 2, or 4, or 8. Let us now choose any $(d+1)$
 number of different linearly dependent normalized states $|{\chi_j}\rangle =
 a_{j0}|0\rangle + a_{j1}|1\rangle + \dots + a_{j(d-1)}|d-1\rangle
 $ (for $j = 1, 2, \dots d+1)$ states, each with real
 coefficients, and such that any $d$ of them are linearly
 independent \cite{ft2}. Then we run our above mentioned argument
 of $1\rightarrow 2$ cloning of these $(d+1)$ number of states.
 And so, having ${\rm log}_2 d$ cbits of classical communication,
 Bob can extract (probabilistically) ${\rm log}_2 (d+1)$ cbits of
 information, by probabilistically distinguishing the $(d+1)$
 number of linearly independent states
 $|\psi_j\rangle\otimes|\psi_j\rangle$  ( for $j = 1, 2,\dots
 d+1$) -- a contradiction. Thus we see that probabilistic
 exact $1 \rightarrow 2$ cloning of any $(d+1)$ number of different
 linearly dependent states of $d$ dimensional Hilbert space (where
 any $d$ among these $(d+1)$ number of states are linearly
 independent), each with real amplitudes, implies probabilistic signalling,
 in the case when $d$ = 2, or  4, or  8.

In conclusion, we have given here an alternative proof of
the result of Hardy and Song \cite{hardy99} for the case of
qubits. Hence no three different states of ${C\!\!\!\!I}^d$
can be probabilistically exactly cloned. For states of
${C\!\!\!\!I}^d$, when $d > 2$, we have alternatively proved
a partial result of ref. \cite{hardy99}, namely, $( d + 1 )$
number of linearly dependent states (of which any $d$ number
of states are linearly independent) taken from a special
ensemble of ${C\!\!\!\!I}^d$, can not be probabilistically
exactly cloned. Although our method does not reproduce the
results of ref. \cite{hardy99}, in full generality, our approach
seems to be comparatively simpler. We loose here the generality of
the argument because of the fact that an arbitrarily given set of
states of ${C\!\!\!\!I}^d$ can not be remotely prepared (not in
asymptotic sense) with  ${\rm log}_2 d$ cbits of classical communication. \\

{\bf Acknowledgement}: The authors would like to thank R. Simon
and P. S. Joag for interesting discussions. Authors are also
thankful to A. K. Pati for useful discussions, in particular, about
the work \cite{zeng}. S. K. acknowledges the support by the
Council of Scientific and Industrial Research, Government of
India, New Delhi. S. K. also acknowledges the support of The
Institute of
Mathematical Sciences, Chennai, for their hospitality, during the work.\\


\begin{thebibliography}{99}
\bibitem{ft} In fact, in order to show that 2 bit of classical
information from Alice to Bob is neccessary in teleporting exactly
an arbitrary qubit from Alice's place to Bob's place,
Bennett et. al., in ref.  \cite{bennett}, have used the same kind
of argument of signalling, as we have used in this paper.
\bibitem{ft1} Such an $U$ does not always exist.
\bibitem{ft2} Such kind of choice of $(d + 1)$ number of states $|\chi_j\rangle$'s can, in general, be made.
\end{thebibliography}
\end{document}